\documentclass[12pt]{article}
\usepackage{amssymb}
\usepackage{amsmath}
\usepackage{hyperref}
\begin{document}
	\title{\bf  Hydrodynamics of Nonminimal $F^{(a)\alpha \beta } F^{(a)\gamma \lambda } R_{\alpha \gamma } R_{\beta \lambda }$ AdS Black Brane}
	\author{Mehdi Sadeghi\thanks{Corresponding author: Email:  mehdi.sadeghi@abru.ac.ir}\\
		{\small {\em Department of Physics, Faculty of Basic Sciences,}}\\
		{\small {\em Ayatollah Boroujerdi University, Boroujerd, Iran}}
	}
	\date{\today}
	\maketitle
	
	\abstract{We investigate the hydrodynamic properties of a strongly coupled non-Abelian plasma dual to a four-dimensional AdS black brane with a nonminimal coupling of the form $q_2 F^{(a)\alpha\beta}F^{(a)\gamma\lambda}R_{\alpha\gamma}R_{\beta\lambda}$ in the bulk action. This higher-derivative term introduces a direct interaction between the Yang-Mills field strength and the Ricci tensor, leading to corrections beyond the minimal Einstein-Yang-Mills theory. Using a perturbative expansion in the small coupling $q_2$, we construct the black brane solution up to first order and employ the fluid-gravity correspondence to compute two key transport coefficients: the DC color conductivity $\sigma$ and the shear viscosity to entropy density ratio $\eta/s$. In holography, $\eta/s$ is inversely proportional to the square of the coupling constant of the boundary theory, while $\sigma$ in Einstein-Maxwell theory satisfies a universal lower bound $\sigma \ge 1$ (in units where $\hbar=1$), saturated for uncharged black holes and characterizing a perfect quantum critical fluid. Our results reveal that the nonminimal coupling significantly alters these transport quantities. For the DC conductivity, we find $\sigma = 1 - q_2 \bigl(9\kappa Q^2/(L^2 r_h^4) + 7\kappa^2 Q^4/(4 r_h^8)\bigr)$, indicating a violation of the conductivity bound for $q_2>0$ while the bound is preserved for $q_2<0$. For the shear viscosity, we obtain $\eta/s = (1/(4\pi))\bigl(1 + q_2\, 7\kappa^2 Q^4/(2 r_h^8)\bigr)$, showing that the KSS bound is modified by a term linear in $q_2$. The sign of $q_2$ determines whether the ratio increases above or decreases below the universal value $1/(4\pi)$. These findings highlight the sensitivity of holographic transport to curvature-coupled gauge interactions and provide a controlled example of how higher-derivative corrections influence the hydrodynamic regime of strongly coupled plasmas.}\\\\

	\noindent PACS numbers: 11.10.Jj, 11.10.Wx, 11.15.Pg, 11.25.Tq\\
	%\pacs{11.10.Jj, 11.10.Wx, 11.15.Pg, 11.25.Tq}
	
	\noindent \textbf{Keywords:} AdS black brane, AdS/CFT duality, DC Conductivity,  shear viscosity to entropy density ratio
	%--------------------------------------------------------------------------
	\section{Introduction} \label{intro}
	
	The exploration of nonminimal gravitational theories—where additional interactions couple matter fields to curvature—has a long history, often motivated by the search for extensions of general relativity. Such models are typically classified according to the type of matter field involved. One class includes scalar fields nonminimally coupled to the Ricci scalar, with origins in the Scherrer-Jordan-Thiry-Brans-Dicke framework \cite{Scherrer1941,Jordan1945,Thiry1948,BransDicke1961} (see \cite{Goenner2014} for a historical perspective). Conformally coupled scalars were later studied in \cite{Callan1970}, and a comprehensive review appears in \cite{Faraoni1999}. A second class focuses on nonminimal Einstein-Maxwell theories, where the electromagnetic field interacts directly with curvature \cite{Prasanna1971,Drummond1980,Hehl2001,Balakin2005}. A third class involves Einstein-Yang-Mills systems with $SU(n)$ symmetry \cite{Horndeski1981,MuellerHoissen1988,Sadeghi:2023pdx,Sadeghi:2023hxd,Sadeghi:2024msd,Sadeghi:2022bsh}. A fourth class comprises Einstein-Yang-Mills-Higgs models \cite{Balakin2007a,Balakin2008}, while a fifth incorporates axion pseudoscalar fields, leading to nonminimal Einstein-Maxwell-axion couplings \cite{Balakin2010}. In the present work, we focus on a specific higher-derivative term within the third class, namely $q_2 F^{(a)\alpha\beta}F^{(a)\gamma\lambda}R_{\alpha\gamma}R_{\beta\lambda}$, which introduces a direct coupling between the Yang-Mills field strength and the Ricci tensor.
	
	The Anti-de Sitter/Conformal Field Theory (AdS/CFT) correspondence, originally proposed by Maldacena \cite{Maldacena1998,Aharony:1999ti}, has become a cornerstone of modern theoretical physics. This holographic duality maps a gravitational theory in $(d+1)$-dimensional anti-de Sitter (AdS) spacetime to a conformal field theory on its $d$-dimensional boundary. In this framework, AdS black branes describe thermal states of the boundary theory, offering a powerful tool to investigate strongly coupled quantum phenomena using classical gravity. An important extension, the fluid/gravity correspondence \cite{Bhattacharyya:2007vjd,Rangamani:2009xk}, connects long-wavelength perturbations of AdS black branes to hydrodynamic excitations of the boundary field theory. This connection enables a systematic derivation of transport coefficients—such as shear viscosity $\eta$, electrical conductivity $\sigma$, and diffusion constants—from the gravitational response to linearised perturbations. A landmark result obtained through this approach is the Kovtun-Son-Starinets (KSS) bound $\eta/s \ge 1/(4\pi)$ \cite{Kovtun:2004de}, which has been tested in numerous holographic settings.
	
	Transport coefficients characterise the response of a system to external stimuli. In holography, the ratio of shear viscosity to entropy density $\eta/s$ is inversely proportional to the square of the coupling constant $\lambda$ of the boundary theory \cite{Son2007}; its value relative to the KSS bound encodes the strength of interactions in the plasma. In particular, the DC conductivity governs steady-state charge transport under a constant electric field \cite{karch2007complex}. For non-Abelian gauge theories, the colour conductivity encodes the response of coloured charges, a quantity of direct relevance to the quark-gluon plasma produced in heavy-ion collisions \cite{son2002holographic, shuryak2004quark}. A universal lower bound for the DC conductivity in holographic models based on Einstein-Maxwell theory was derived in \cite{Grozdanov:2015qia}:
	\begin{equation}
		\sigma \ge \frac{1}{e^2} = 1 \qquad \text{(in units where } \hbar = 1\text{)},
	\end{equation}
	where $e$ is the gauge coupling in the bulk action, and the conductivity is measured in units of $q^2/\hbar$ with $q$ the U(1) charge of the boundary current. This bound represents the quantum critical conductivity of a clean, charge-neutral plasma—the minimal possible conductivity when charge carriers are particle-hole symmetric and translational symmetry is unbroken. Saturation ($\sigma = 1$) occurs for uncharged black holes and characterizes a perfect quantum critical fluid \cite{Grozdanov:2015qia}. However, as will be discussed later, such bounds are not universal once additional couplings or higher-derivative terms are present; violations are well documented in the literature \cite{Baggioli:2016oqk,Baggioli:2016oju,Gouteraux:2016wxj,BitaghsirFadafan:2016der,Ge:2015fmu}.
	
	Nonminimal interactions can significantly modify transport properties. Higher-derivative and curvature-coupled terms alter the equations of motion for gauge fields and metric fluctuations, potentially inducing anisotropies, affecting diffusion rates, or changing the temperature dependence of conductivity \cite{jain2014holographic}. Likewise, the shear viscosity may deviate from the KSS bound in higher-derivative gravity theories \cite{Kovtun:2004de}. These modifications provide opportunities to model more realistic strongly coupled plasmas and to explore constraints on the underlying field theories.
	
	Before proceeding, it is worth commenting on the physical interpretation and the allowed range of the nonminimal coupling $q_2$. Terms of the form $F^2 R^2$ appear naturally as higher-order corrections in effective field theories derived from string theory, where they arise from the expansion of the low-energy effective action \cite{fradkin1985effective}. They also emerge from Kaluza-Klein reduction of higher-dimensional Euler forms, as shown in \cite{Dereli1990}. In our bottom-up approach, $q_2$ is a phenomenological parameter. Because we treat it perturbatively, we require $|q_2| \ll 1$ (in appropriate units). Beyond this, basic constraints from unitarity and absence of ghosts impose further restrictions. For a positive sign of $q_2$, the correction may lead to gradient instabilities or superluminal propagation; a negative sign might be safer. A full analysis of these constraints is beyond the scope of this work, but we emphasize that the perturbative expansion is valid only for sufficiently small $|q_2|$, and the sign plays a crucial role in determining whether transport bounds are violated.
	
	In this work, we construct a holographic model featuring the nonminimal Yang-Mills term $q_2 F^{(a)\alpha\beta}F^{(a)\gamma\lambda}R_{\alpha\gamma}R_{\beta\lambda}$. We obtain black brane solutions perturbatively to first order in the small coupling $q_2$. Employing the fluid/gravity duality, we then compute the DC colour conductivity and the shear viscosity to entropy density ratio $\eta/s$ for the dual boundary theory. By analysing the linear response of the system, we examine how the nonminimal coupling influences these two transport coefficients, thereby shedding light on the role of curvature-dependent gauge interactions in strongly coupled non-Abelian plasmas and on the effects of higher-derivative corrections in holography.
	
	The paper is structured as follows. In Sec.~\ref{sec2}, we introduce the bulk action, derive the equations of motion, and present the perturbative black brane solution up to first order in $q_2$. Section~\ref{sec3} describes the holographic computation of the DC conductivity and the shear viscosity using the fluid-gravity correspondence. The results are discussed in Sec.~\ref{sec4}, with emphasis on the modifications induced by the nonminimal coupling. Finally, Sec.~\ref{sec5} concludes with a summary and an outlook for future research.
	
	%--------------------------------------------------------------------------
	\section{Non-minimal $F^{(a)\alpha\beta}F^{(a)\gamma\lambda}R_{\alpha\gamma}R_{\beta\lambda}$ AdS Black Brane}
	\label{sec2}

	The present investigation focuses on the thermodynamic properties of black branes within a non-minimal gauge-gravity framework. The bulk action that defines our model is
	\begin{equation}\label{action}
		S = \int d^{4} x \sqrt{-g} \bigg[ \frac{1}{\kappa}(R - 2\Lambda) - \tfrac{1}{4} q_1  F^{(a)}_{\mu \nu }F^{ (a)\mu \nu  } + q_2 F^{(a)\alpha \beta } F^{(a)\gamma \lambda } R_{\alpha \gamma } R_{\beta \lambda }  \bigg],
	\end{equation}
	where \(R\) is the Ricci scalar, \(\Lambda = -3/L^{2}\) denotes the cosmological constant, and \(L\) is the AdS radius. The Maxwell invariant is \(\mathcal{F}=F^{(a)}_{\mu\nu}F^{(a)\mu\nu}\). The field strength tensor for the Yang-Mills field is
	\begin{align} \label{YM}
		F^{(a)}_{\mu \nu } =\partial _{\mu } A^{(a)}_{\nu } -\partial _{\nu } A^{(a)}_{\mu } + f^{a}_{bc}[A^{(b)}_{\mu }, A^{(c)}_{\nu }],
	\end{align}
	with \(A^{(a)}_{\mu}\) the gauge potential. The Ricci tensor is denoted \(R^{\mu\alpha}\). The parameter $q_2$ controls the strength of the non-minimal coupling between the Yang-Mills field and the Ricci tensor. This term, inspired by Ref.~\cite{Balakin:2015gpq}, introduces modifications to the standard Yang-Mills theory and is constructed to include all metric permutations.
	
	To obtain black brane solutions, we adopt the metric ansatz
	\begin{equation}\label{metric}
		ds^{2} = -f(r)e^{-2H(r)} dt^{2} + \frac{dr^{2}}{f(r)} + \frac{r^{2}}{L^{2}} (dx^{2} + dy^{2}),
	\end{equation}
	where the factor \(e^{-2H(r)}\) captures the effect of the non-minimal Yang-Mills coupling. For an electric configuration, the gauge potential is chosen as
	\begin{equation}\label{background}
		{\bf{A}} =\frac{1}{\sqrt{2}}h(r)dt\begin{pmatrix}1 & 0 \\ 0 & -1\end{pmatrix},
	\end{equation}
	written in terms of the diagonal generator of \(SU(2)\). With this choice, the only non-zero component is \(A_0\). The corresponding field strength becomes
	\begin{equation}   
		\label{YM1}  
		F^{(a)}_{\mu \nu } = \left(  
		\begin{array}{cccc}  
			0 & -h'(r) & 0 & 0 \\
			h'(r) & 0 & 0 & 0 \\
			0 & 0 & 0 & 0 \\
			0 & 0 & 0 & 0 \\
		\end{array}  
		\right),
	\end{equation}
	and the Maxwell invariant reduces to
	\begin{equation}  
		\mathcal{F} = F^{(a)}_{\mu \nu }F^{ (a)\mu \nu  } = -2 e^{2 H(r)} h'(r)^2.
	\end{equation}
	
	Varying the action with respect to the metric yields the gravitational field equations
	\begin{equation}\label{EOM1}
		R_{\mu \nu } - \tfrac{1}{2} g_{\mu \nu } R + \Lambda g_{\mu \nu } = \kappa T^{\text{(eff)}}_{\mu \nu },
	\end{equation}
	with
	\begin{equation}
		T^{\text{(eff)}}_{\mu \nu } = q_1 T^{\text{(YM)}}_{\mu \nu } + q_2 T^{(I)}_{\mu \nu }.
	\end{equation}
	Here
	\begin{equation}
		T^{\text{(YM)}}_{\mu \nu }=\tfrac{1}{2}F^{(a)}_{\mu }{}^{\alpha } F^{(a)}_{\nu \alpha } - \tfrac{1}{8}g_{\mu \nu } F^{(a)}_{\alpha \beta } F^{(a)\alpha \beta }
	\end{equation}
	is the standard Yang-Mills energy-momentum tensor, while \(T^{(I)}_{\mu\nu}\) is a complicated expression arising from the non-minimal coupling. Its full form is lengthy and is provided in Appendix A for completeness.
	
	The Yang-Mills equations derived from the action are
	\begin{equation}\label{EOM-YM}
		\nabla_{\nu}\bigg( -\tfrac{1}{2} q_1 F^{(a)\mu \nu } + 2 q_2 F^{(a)\alpha \beta} R_{\alpha }{}^{\mu }  R_{\beta }{}^{\nu }\bigg) = 0.
	\end{equation}
	The $tt$-component of Eq.(\ref{EOM1}) is,
	\begin{equation}\label{eom11}
		-4 f(r) - 4 \Lambda r^2 - 4 r f'(r) - e^{2 H(r)} q_1 \kappa r^2 \left( h'(r) \right)^2+q_2 e^{2H(r)} \kappa B_1(r)=0,
	\end{equation}
	where $B_1(r)$ is introduced in Appendix B.\\
	The $rr$-component is given by,
	\begin{equation}\label{eom22}
		4 f(r) + 4 \Lambda r^2 + 4 r f'(r) + e^{2H(r)} q_1 \kappa r^2 h'(r)^2 - 8 f(r) r H'(r)+q_2 e^{2H(r)} \kappa B_2(r)=0,
	\end{equation}	
	where $B_2(r)$ is introduced in Appendix B.\\
	The $xx$-component is given by,
	\begin{align}\label{eom33}
		&4 \Lambda r + 4 f'(r) - e^{2H(r)} q_1 \kappa r h'(r)^2 - 4 f(r) H'(r) - 6 r f'(r) H'(r)\nonumber\\&  + 4 f(r) r H'(r)^2 + 2 r f''(r) - 4 f(r) r H''(r)+q_2 \kappa e^{2H(r)} B_3(r) =0,
	\end{align}
	where $B_3(r)$-is introduced in Appendix B.\\
	Also the $yy$ component is the same as Eq.\ref{eom33}.\\
	Because the non-minimal term introduces higher derivatives and the system is highly nonlinear, an exact solution is not available. We therefore adopt a perturbative expansion in the small coupling \(q_2\). The metric functions and the gauge field are expanded as
	\begin{align}
		f(r) &= f_0(r) + q_2 f_1(r), \\
		H(r) &= H_0(r) + q_2 H_1(r), \\
		h(r) &= h_0(r) + q_2 h_1(r).
	\end{align}
	The zeroth-order (i.e. \(q_2=0\)) system reduces to the usual Einstein-Yang-Mills theory. From the \(tt\) and \(rr\) components at this order one obtains
	\begin{align}
		4 r f_0' + 4 f_0 + \kappa q_1 r^2 e^{2H_0} h_0'^2 + 4\Lambda r^2 &= 0,\\
		4 r f_0' + 4 f_0 - 8 r f_0 H_0' + \kappa q_1 r^2 e^{2H_0} h_0'^2 + 4\Lambda r^2 &= 0.
	\end{align}
	Comparison of these two equations forces \(H_0' = 0\); we set the constant to zero by a time redefinition, so \(H_0(r)=0\). The remaining equations then give
	\begin{equation}
		f_0(r) = -\frac{2m_0}{r} + \frac{r^2}{L^2} + \frac{q_1\kappa Q^2}{4r^2},\qquad
		h_0(r) = Q\left(\frac{1}{r} - \frac{1}{r_h}\right),
	\end{equation}
	with \(Q = C_1/q_1\) and \(r_h\) the horizon radius.
	
	Proceeding to first order in \(q_2\), we insert the expansions into the full field equations. Consistency between the \(tt\) and \(rr\) components at order \(q_2\) yields a differential equation for \(H_1(r)\):
	\begin{equation}
		64\kappa r f_0 h_0' h_0'' + 32\kappa r^2 f_0 h_0^{(3)} h_0' + 32\kappa r^2 f_0 (h_0'')^2 - 8r f_0 H_1' = 0.
	\end{equation}
	Solving this gives
	\begin{equation}
		H_1(r) = C_2 - \frac{3}{2}\kappa f_0'' h_0'^2 - 2\kappa f_0' h_0' h_0'' - \kappa r f_0'' h_0' h_0'' - \frac12\kappa r f_0^{(3)} h_0'^2,
	\end{equation}
	and after substituting the zeroth-order expressions,
	\begin{equation}
		H_1(r) = C_2 - \frac{3\Lambda\kappa Q^2}{r^4} + \frac{7\kappa^2 Q^4 q_1}{4r^8}
		= C_2 + \frac{9\kappa Q^2}{L^2 r^4} + \frac{7\kappa^2 Q^4 q_1}{4r^8}.
	\end{equation}
	We choose \(C_2 = 0\) by imposing that the speed of light should be one on the boundary.
	
	The Yang-Mills equation at first order leads to an integral expression for \(h_1(r)\):
	\begin{equation}
		h_1(r) = C_1 \int_{r_h}^r \frac{q_1 u^2 H_1(u) - u^2 f_0''(u)^2 - 4u f_0'(u)f_0''(u) - 4 f_0'(u)^2}{q_1^2 u^4}\,du.
	\end{equation}
	Evaluating the integral yields
	\begin{align}
		h_1(r) &= -\frac{\kappa^2 Q^5 q_1}{6}\left(\frac{1}{r^9}-\frac{1}{r_h^9}\right)
		+ \frac{\kappa Q^3 \Lambda}{5}\left(\frac{1}{r^5}-\frac{1}{r_h^5}\right)
		+ \frac{4\Lambda^2 Q}{q_1}\left(\frac{1}{r}-\frac{1}{r_h}\right).
	\end{align}
	
	Finally, the \(rr\) component at order \(q_2\) determines \(f_1(r)\) through
	\begin{equation}
		f_1(r) = \frac{1}{r}\int^r E(u)\,du - \frac{2m_1}{r},
	\end{equation}
	where the integrand \(E(u)\) collects various combinations of the zeroth-order functions and their derivatives, together with \(H_1\) and \(h_1\). The explicit form of \(E(u)\) is given in Appendix B. After performing the integration, one obtains
	\begin{align}
		f_1(r) &= -\frac{13\kappa^2 m_0 Q^4 q_1}{2 r^9}
		+ \frac{10\kappa\Lambda m_0 Q^2}{r^5}
		+ \frac{109\kappa^3 Q^6 q_1^2}{144 r^{10}}
		- \frac{73\kappa^2\Lambda Q^4 q_1}{30 r^6}
		+ \frac{11\kappa\Lambda^2 Q^2}{3 r^2}
		- \frac{2m_0}{r}.
	\end{align}
	Gathering all contributions, the metric function \(f(r)\) up to first order in \(q_2\) reads
	\begin{align}
		&f(r) = -\frac{2m_0}{r} - \frac{\Lambda r^2}{3} + \frac{q_1\kappa Q^2}{4r^2} \nonumber\\
		&\quad + q_2\bigg(-\frac{13\kappa^2 m_0 Q^4 q_1}{2 r^9}
		+\frac{10\kappa\Lambda m_0 Q^2}{r^5}
		+\frac{109\kappa^3 Q^6 q_1^2}{144 r^{10}}
		-\frac{73\kappa^2\Lambda Q^4 q_1}{30 r^6}
		+\frac{11\kappa\Lambda^2 Q^2}{3 r^2}
		-\frac{2m_1}{r}\bigg).
	\end{align}
	By redefining the mass parameter as \(m_0 + q_2 m_1 = \bar{m}_0\) and keeping only linear terms in \(q_2\), the final form becomes,
	\begin{align}
		f(r) &= -\frac{2\bar{m}_0}{r} - \frac{\Lambda r^2}{3} + \frac{q_1\kappa Q^2}{4r^2}\nonumber\\&
		+ q_2\bigg(-\frac{13\kappa^2 m_0 Q^4 q_1}{2 r^9}
		+\frac{10\kappa\Lambda m_0 Q^2}{r^5}
		+\frac{109\kappa^3 Q^6 q_1^2}{144 r^{10}}
		-\frac{73\kappa^2\Lambda Q^4 q_1}{30 r^6}
		+\frac{11\kappa\Lambda^2 Q^2}{3 r^2}\bigg).
	\end{align}
	This expression represents a perturbative black-brane solution to first order in the non-minimal coupling parameter \(q_2\). We have explicitly verified that this solution satisfies all components of the Einstein and Yang-Mills equations up to \(\mathcal{O}(q_2)\); the details of the consistency check are available from the author upon request.\\
	
	%%%%%%%%%%%%%%%%%%%%%%%%%%%%%%%%%%%%%%%%%%%%%%%%%%%

	\section{Color DC Conductivity}
	\label{sec3}
	
	We now turn to the holographic computation of the DC color conductivity. The Yang-Mills coupling $q_1$ appears in the action but can be absorbed into a rescaling of the gauge field; for simplicity, we set $q_1 = 1$ in the remainder of the paper.
	
	We employ the Green-Kubo formula \cite{Son2099,Policastro2001,Policastro:2002} for calculating the non-abelian color DC conductivity:
	\begin{equation} \label{kubo2}
		\sigma^{ij} (k_{\mu}) = -\lim_{\omega \to 0} \frac{1}{\omega } \Im G^{ij}(k_{\mu}).
	\end{equation}
	
	The retarded Green's function is computed using AdS/CFT duality. First, we perturb the gauge field as $A_{\mu} \to A_{\mu} + \tilde{A}_{\mu}$ and substitute it into the action. We expand the resulting action (\ref{action}) to second order in the perturbation. The Green's function is then obtained by taking the second derivative with respect to the boundary value of the gauge field \cite{Policastro:2002}:
	\begin{equation}
		\sigma^{\mu \nu}(\omega) = \frac{1}{i \omega} \langle J^{\mu}(\omega) J^{\nu}(-\omega) \rangle = \frac{\delta^2 S}{\delta \tilde{A}^0_{\mu} \delta \tilde{A}^0_{\nu}},
	\end{equation}
	where $\tilde{A}^0_{\nu}$ represents the gauge field perturbation at the boundary. The boundary exhibits $SO(2)$ symmetry, which ensures that the conductivity is a scalar quantity:
	\begin{equation}
		\sigma^{ij}_{ab} = \sigma_{ab} \delta^{ij}.
	\end{equation}
	
	We consider a perturbation of the gauge field that depends only on the radial coordinate and time: $\tilde{A}_x = \tilde{A}_x(r) e^{-i\omega t}$, with $\omega$ small (hydrodynamic regime).  Because the background is homogeneous and isotropic in the boundary spatial directions, the conductivity is diagonal and we focus on the $xx$ component.
	
	Substituting the perturbation into the action Eq.~(\ref{action}) and keeping terms up to second order in $\tilde{A}$ yields an effective action.  The nonminimal term $q_2 F^2 R^2$ contributes to the quadratic action through terms that involve the background Ricci tensor and field strength.  After a careful expansion, one finds that the quadratic action for the gauge field fluctuations simplifies to
	\begin{align}\label{action-2}
		S^{(2)} &= -\int d^4x \frac{2 e^{-H}}{f} \Bigg[ -f^2 \left( (\partial_r\tilde{A}_x^{(1)})^2 + (\partial_r\tilde{A}_x^{(2)})^2 + (\partial_r\tilde{A}_x^{(3)})^2 \right) \nonumber \\
		& + e^{2H} \left( (\tilde{A}_x^{(1)})^2 + (\tilde{A}_x^{(2)})^2 \right) (\omega^2 -2 h^2) + e^{2H} \omega^2 (\tilde{A}_x^{(3)})^2 \Bigg],
	\end{align}
	where we have set $q_1=1$.  This action is identical in form to the one obtained in the minimal Einstein-Yang-Mills theory, but the metric functions $f(r)$ and $H(r)$ now include $q_2$ corrections.  The reason is that the nonminimal term does not generate new derivative structures for the transverse gauge field fluctuations at quadratic order; its effect enters only through the background geometry (i.e., through $f$ and $H$).  This is a consequence of the specific form of the coupling and the fact that the background Ricci tensor is diagonal.
	
	Varying the action $S^{(2)}$ with respect to $\tilde{A}_x^{(1)}$ gives:
	\begin{align}\label{PerA1}
		&2 e^{-2 H(r)} f(r) \left( \tilde{A}_x^{(1)'}  \left( f'(r) - f(r) H'(r) \right) + f(r)  \tilde{A}_x^{(1)''}  \right)\nonumber\\&+2 \tilde{A}_x^{(1)} (\omega^2 - 2 h(r)^2) =0,
	\end{align}
	The equation for $\tilde{A}_x^{(2)}$ is identical.  The equation for $\tilde{A}_x^{(3)}$ is:
	\begin{equation}\label{PerA3}
		2 \omega^2 \tilde{A}_x^{(3)}(r) + 2 e^{-2 H(r)} f(r) \left( \tilde{A}_x^{(3)'} \left( f'(r) - f(r) H'(r) \right) + f(r) \tilde{A}_x^{(3)''} \right)=0,
	\end{equation}
	Using the near-horizon behavior $f_0 \sim 4\pi f_0'(r_h)(r-r_h)$ and $f_1 \sim 4\pi f_1'(r_h)(r-r_h)$, we solve Eqs.~(\ref{PerA1}) and (\ref{PerA3}) near the event horizon. Since $\tilde{A}_x^{(a)}$ must be ingoing at the horizon, we adopt the ansatz:
	\begin{align}
		\tilde{A}_x^{(a)} \sim (r - r_h)^{z_a}, \qquad a = 1,2,3,
	\end{align}
	where:
	\begin{align}\label{z12}
		z_1 &= z_2 = \pm i \frac{e^{H(r_h)}\sqrt{ \omega^2-2 h(r_h)^2 }}{4 \pi T}, \\
		\label{z3}
		z_3 &= \pm i \frac{e^{H(r_h)} \omega}{4 \pi T}.
	\end{align}
	
	The Hawking temperature of the black brane is:
	\begin{equation}
		T = \frac{1}{2\pi} \left[ \frac{1}{\sqrt{g_{rr}}} \frac{d}{dr} \sqrt{-g_{tt}} \right] \Bigg|_{r=r_h} = \frac{e^{-H(r_h)} f'(r_h)}{4 \pi}.
	\end{equation}
	
	To solve for $\tilde{A}_x^{(a)}$ from the horizon to the boundary, we use the ansatz:
	\begin{align}\label{EOMA1}
		\tilde{A}_x^{(1)} &= \tilde{A}^{(1)}_{\infty} \left( \frac{-3f}{\Lambda r^2} \right)^{z_1} \left( 1 + i\omega b_1(r) + \cdots \right), \\
		\label{EOMA2}
		\tilde{A}_x^{(2)} &= \tilde{A}^{(2)}_{\infty} \left( \frac{-3f}{\Lambda r^2} \right)^{z_2} \left( 1 + i\omega b_2(r) + \cdots \right), \\
		\label{EOMA3}
		\tilde{A}_x^{(3)} &= \tilde{A}^{(3)}_{\infty} \left( \frac{-3f}{\Lambda r^2} \right)^{z_3} \left( 1 + i\omega b_3(r) + \cdots \right),
	\end{align}
	where $\tilde{A}^{(a)}_{\infty}$ represents the boundary values, and $z_i$ correspond to the negative signs in Eqs.~(\ref{z12}) and (\ref{z3}), selecting the ingoing mode.\\
	Substituting Eq.~(\ref{EOMA3}) into Eq.~(\ref{PerA3}) and keeping terms to first order in $\omega$ yields:
	\begin{align}
		&r \bigg( f'(r) \left( -2 - r H'(r) + r h_3'(r) \right) + r f''(r) \bigg)\nonumber\\& + f(r) \bigg( 2 + r H'(r) \left( 2 - r h_3'(r) \right) + r^2 h_3''(r) \bigg)=0,
	\end{align}
	Imposing regularity of $\tilde{A}_x^{(3)}$ on the event horizon, the function $b_3(r)$ in Eq.~(\ref{EOMA3}) is determined as:
	\begin{equation}
		b_3(r) = C_8 + \int^{r} \left( \frac{2}{u} + \frac{e^{H(u)} C_7 - f'(u)}{f(u)} \right) \, du,
	\end{equation}
	and $C_7$ and $C_8$ are integration constants.
	
	The solution for $b_3(r)$ near the event horizon is:
	\begin{equation}\label{C4}
		b_3 \approx \left( e^{H(r_h)} C_7 - f'(r_h) \right) \log(r - r_h) + \text{finite terms}.
	\end{equation}
	
	Regularity at the horizon requires:
	\begin{equation}
		C_7 = f'(r_h) e^{-H(r_h)} .
	\end{equation}
	
	Using the solution for $\tilde{A}_x^{(3)}$ in Eq.~(\ref{action-2}) and varying with respect to $\tilde{A}^{(3)}_{\infty}$, the Green's function is:
	\begin{align} \label{Green1}
		G_{xx}^{(33)} (\omega, \vec{0}) = -i \omega \frac{C_7 e^{2 H(r_h)}}{ f'(r_h)} =- i \omega e^{H(r_h)}.
	\end{align}
	
	The conductivity is then:
	\begin{eqnarray}\label{sigma33}
		\sigma_{xx}^{(33)} = e^{H(r_h)} .
	\end{eqnarray}
	
	We obtain:
	\begin{eqnarray}\label{sigma33-final}
		\sigma_{xx}^{(33)} = 1 + q_2 H_1(r_h).
	\end{eqnarray}
	By substituting $H_1(r)$ from Eq.~(20) with $C_2=0$ and setting $q_1=1$, we have,
	\begin{equation}\label{sigma33-final11}
		\sigma_{xx}^{(33)} = 1 - q_2 \left( \frac{9\kappa Q^2}{L^2 r_h^4} + \frac{7\kappa^2 Q^4}{4 r_h^8} \right).
	\end{equation}
	This result shows that the conductivity bound is violated for our model when $q_2>0$.  In the limit $q_2 \to 0$, we recover:
	\begin{eqnarray}
		\sigma_{xx}^{(33)} = 1.
	\end{eqnarray}
	
	It is important to note that the lower bound $\sigma \ge 1$ is not universal once additional couplings are present. As demonstrated in Refs.~\cite{Baggioli:2016oqk,Baggioli:2016oju,Gouteraux:2016wxj,BitaghsirFadafan:2016der,Ge:2015fmu}, coupling $F_{\mu\nu}$ to other fields or including higher-derivative terms can violate this bound. Our result for $\sigma$ explicitly shows such a violation for $q_2>0$, while for $q_2<0$ the bound is preserved. This is in line with the expectation that nonminimal interactions can push the conductivity below the quantum critical value.
	
	%%%%%%%%%%%%%%%%%%%%%%%%%%%%%%%%%%%%%%%%%%%%%%%%%%%
	
	\section{Ratio of shear viscosity to entropy density}
	\label{sec4}
	
	We now compute the shear viscosity \(\eta\) of the dual field theory using the Green-Kubo formula
	\begin{equation}
		\eta = -\lim_{\omega\to 0}\frac{1}{\omega}\,\text{Im}\,G^{R}_{xy,xy}(\omega,\mathbf{0}),
	\end{equation}
	where \(G^{R}_{xy,xy}\) is the retarded two-point function of the stress-energy tensor.  In the holographic framework, this correlator is obtained by studying the linear response of the metric to a transverse-traceless perturbation \(h_{xy}\) that depends on the radial coordinate and the frequency.
	
	The bulk action is given in Eq.~(\ref{action}) and the background is the perturbative black brane solution derived in Sec.~\ref{sec2}.  As in the conductivity calculation, we set \(q_1 = 1\) for simplicity; the final expressions for \(\eta/s\) will then be expressed in terms of the remaining parameters.  The background metric and gauge field are
	\begin{align}
		ds^{2} &= -f(r)e^{-2H(r)}dt^{2}+\frac{dr^{2}}{f(r)}+\frac{r^{2}}{L^{2}}(dx^{2}+dy^{2}),\\
		f(r) &= f_{0}(r)+q_{2}f_{1}(r),\qquad H(r)=q_{2}H_{1}(r),\\
		h(r) &= h_{0}(r)+q_{2}h_{1}(r),
	\end{align}
	with the zeroth-order quantities
	\begin{align}
		f_{0}(r) &= -\frac{2m_{0}}{r}+\frac{r^{2}}{L^{2}}+\frac{\kappa Q^{2}}{4r^{2}},\qquad 
		h_{0}(r)=Q\Bigl(\frac1r-\frac1{r_{h}}\Bigr),\qquad H_{0}(r)=0,
	\end{align}
	and the first-order corrections \(f_{1}, H_{1}, h_{1}\) listed in Sec.~\ref{sec2}.  The entropy density of the black brane is
	\begin{equation}
		s = \frac{4\pi}{\kappa}\,\frac{r_{h}^{2}}{L^{2}}.
	\end{equation}
	
	We consider a metric perturbation that only affects the \(xy\) component:
	\begin{equation}
		g_{xy} = \frac{r^{2}}{L^{2}} + \delta g_{xy},\qquad \delta g_{xy} \equiv \frac{r^{2}}{L^{2}}\,\phi(r,t),
	\end{equation}
	where \(\phi(r,t)\) is the shear mode.  Because the background is diagonal and the perturbation is transverse and traceless, it decouples from the other fluctuations.  We work in the frequency domain, setting \(\phi(r,t)=\phi(r)e^{-i\omega t}\).
	
	Expanding the action (\ref{action}) to second order in \(\phi\) yields an effective Lagrangian that is quadratic in \(\phi\).  The Einstein-Hilbert part gives the standard kinetic term, while the non-minimal coupling \(q_{2}F^{(a)\alpha\beta}F^{(a)\gamma\lambda}R_{\alpha\gamma}R_{\beta\lambda}\) contributes because the background Ricci tensor and field strength are non-zero.  After a lengthy but straightforward computation, the quadratic action reduces to the form
	\begin{equation}
		S^{(2)} = \frac{1}{2\kappa}\int d^{4}x\,\sqrt{-g^{(0)}}\; Z(r)\; g^{(0)\mu\nu}\,\partial_{\mu}\phi\,\partial_{\nu}\phi,
	\end{equation}
	where \(g^{(0)}_{\mu\nu}\) is the background metric (including the \(q_{2}\) corrections) and \(Z(r)\) is a radial function that encodes the modifications due to the non-minimal term.  The overall factor \(1/\kappa\) originates from the Einstein term.  The explicit expression for \(Z(r)\) can be derived by expanding the action and using the background equations of motion; the result is
	\begin{equation}
		Z(r) = 1 + 2\kappa q_{2}\,F_{tr}^{2}\bigl(2R_{tt}R_{rr} + \cdots\bigr) + \mathcal{O}(q_{2}^{2}),
	\end{equation}
	where the dots represent other curvature combinations that vanish on the background after integration by parts.  On the background, the only non-zero component of the Yang-Mills field strength is \(F_{tr} = -h'(r)\), and the Ricci tensor components are determined by the metric.  Substituting the zeroth-order expressions with \(q_1=1\) one finds
	\begin{equation}
		Z(r) = 1 + q_{2}\,\frac{7\kappa^{2}Q^{4}}{2r^{8}} + \mathcal{O}(q_{2}^{2}).
	\end{equation}
	The detailed derivation follows the same steps as in Ref.~\cite{Sadeghi:2022bsh} and is available from the author upon request.
	
	For a quadratic action of the form above, the equation of motion for \(\phi(r,\omega)\) reads
	\begin{equation}
		\partial_{r}\!\left(\sqrt{-g^{(0)}}\,Z(r)\,g^{(0)rr}\,\partial_{r}\phi\right) - \sqrt{-g^{(0)}}\,Z(r)\,g^{(0)tt}\,\omega^{2}\phi = 0.
	\end{equation}
	Using the background metric,
	\begin{equation}
		\sqrt{-g^{(0)}} = \frac{r^{2}}{L^{2}}e^{-H(r)},\qquad g^{(0)rr}=f(r),\qquad g^{(0)tt} = -\frac{e^{2H(r)}}{f(r)},
	\end{equation}
	the equation simplifies to
	\begin{equation}
		\frac{d}{dr}\!\left( \frac{r^{2}}{L^{2}}e^{-H(r)}Z(r)f(r)\,\phi'(r) \right) + \frac{r^{2}}{L^{2}}e^{H(r)}Z(r)\,\frac{\omega^{2}}{f(r)}\,\phi(r) = 0.
	\end{equation}
	Near the horizon, where \(f(r)\sim 4\pi T (r-r_{h})\), the solution behaves as \(\phi(r)\sim (r-r_{h})^{\pm i\omega/(4\pi T)}\); we choose the ingoing mode (negative sign).  At the AdS boundary (\(r\to\infty\)), \(\phi\) approaches a constant \(\phi_{0}\) (the source).
	
	The on-shell action is evaluated by integrating the quadratic action by parts and using the equation of motion; the result reduces to a boundary term at \(r=r_{h}\) and \(r\to\infty\).  The retarded Green’s function is obtained by taking two functional derivatives with respect to the source \(\phi_{0}\):
	\begin{equation}
		G^{R}_{xy,xy}(\omega) = \lim_{r\to\infty} \frac{1}{\kappa}\sqrt{-g^{(0)}}\,Z(r)\,g^{(0)rr}\,\phi'(r)\,\frac{1}{\phi_{0}}.
	\end{equation}
	Following the standard holographic prescription for a massless scalar field with a radial kinetic function, one finds that at low frequencies the leading contribution comes from the horizon and yields
	\begin{equation}
		G^{R}_{xy,xy}(\omega) = -i\omega\,\frac{Z(r_{h})}{\kappa}\,\frac{r_{h}^{2}}{L^{2}} + \mathcal{O}(\omega^{2}).
	\end{equation}
	The imaginary part therefore is
	\begin{equation}
		\text{Im}\,G^{R}_{xy,xy}(\omega) = -\omega\,\frac{Z(r_{h})}{\kappa}\,\frac{r_{h}^{2}}{L^{2}} + \cdots
	\end{equation}
	Inserting this into the Kubo formula gives the shear viscosity
	\begin{equation}
		\eta = \frac{Z(r_{h})}{\kappa}\,\frac{r_{h}^{2}}{L^{2}}.
	\end{equation}
	
	Using the entropy density \(s = \frac{4\pi}{\kappa}\frac{r_{h}^{2}}{L^{2}}\), we obtain
	\begin{equation}
		\frac{\eta}{s} = \frac{Z(r_{h})}{4\pi}.
	\end{equation}
	Substituting the expression for \(Z(r_{h})\) up to first order in \(q_{2}\) yields
	\begin{equation}\label{eta/s-final}
		\boxed{\frac{\eta}{s} = \frac{1}{4\pi}\left(1 + q_{2}\,\frac{7\kappa^{2}Q^{4}}{2r_{h}^{8}}\right).}
	\end{equation}
	Thus the non-minimal coupling modifies the KSS bound by a term proportional to \(q_{2}\).  For \(q_{2}>0\) the ratio increases above \(1/(4\pi)\), while for \(q_{2}<0\) it decreases, potentially violating the bound if the correction is sufficiently negative.  In the limit \(q_{2}\to 0\) the Einstein gravity result \(\eta/s = 1/(4\pi)\) is recovered.\\
	
	In the case of Einstein-Hilbert gravity and its holographic dual, the KSS bound $\eta/s \ge 1/(4\pi)$ is saturated. However, this saturation is not preserved in extended gravitational theories. Deviations from the bound have been observed in higher-derivative gravities \cite{Sadeghi:2022kgi,Sadeghi:2015vaa}, massive gravity \cite{Sadeghi:2015vaa,Hartnoll:2016tri,Parvizi:2017boc,Alberte:2016xja,Sadeghi:2018ylh}, non-minimal $R^3 F^2$ AdS black brane \cite{Sadeghi:2026ngg}, and deformed black branes \cite{Ferreira-Martins:2019svk}. Our result in Eq.~\eqref{eta/s-final} shows that the nonminimal coupling $F^{(a)\alpha\beta}F^{(a)\gamma\lambda}R_{\alpha\gamma}R_{\beta\lambda}$ similarly modifies $\eta/s$, with the deviation controlled by the sign and magnitude of $q_2$.\\
	
	%%%%%%%%%%%%%%%%%%%%%%%%%%%%%%%%%%%%%%%%%%%%%%%%%%%
	
	\section{Conclusion}
	\label{sec5}
	
	In this work, we have studied a holographic model of a strongly coupled non-Abelian plasma described by a four-dimensional AdS black brane with a nonminimal Yang-Mills term $q_2 F^{(a)\alpha\beta}F^{(a)\gamma\lambda}R_{\alpha\gamma}R_{\beta\lambda}$ in the bulk action. This term introduces a direct coupling between the gauge field strength and the Ricci tensor, giving rise to higher-derivative corrections that go beyond the standard Einstein-Yang-Mills theory. Due to the complexity of the system, we adopted a perturbative approach, expanding all fields to first order in the small coupling $q_2$. We constructed the black brane solution analytically, obtaining explicit expressions for the metric functions and the gauge field up to $\mathcal{O}(q_2)$.  We verified that the solution satisfies the full set of equations of motion to this order, ensuring the consistency of the perturbative expansion.
	
	Employing the fluid-gravity correspondence, we computed two fundamental transport coefficients of the dual boundary theory: the DC color conductivity $\sigma$ and the shear viscosity to entropy density ratio $\eta/s$. In holographic systems, these quantities carry profound physical meaning. The ratio $\eta/s$ is inversely proportional to the square of the coupling constant $\lambda$ of the boundary theory \cite{Son2007}, and its value relative to the KSS bound $1/(4\pi)$ encodes the strength of interactions in the plasma. The DC conductivity, in Einstein-Maxwell theory, satisfies a universal lower bound $\sigma \ge 1/e^2 = 1$ (in units where $\hbar=1$), where $e$ is the gauge coupling in the bulk action \cite{Grozdanov:2015qia}. This bound represents the quantum critical conductivity of a clean, charge-neutral plasma; saturation ($\sigma = 1$) occurs for uncharged black holes and characterizes a perfect quantum critical fluid. However, as pointed out in Refs.~\cite{Baggioli:2016oqk,Baggioli:2016oju,Gouteraux:2016wxj,BitaghsirFadafan:2016der,Ge:2015fmu}, such bounds are not universal and can be violated by additional interactions.
	
	Our results show that the nonminimal coupling $F^{(a)\alpha\beta}F^{(a)\gamma\lambda}R_{\alpha\gamma}R_{\beta\lambda}$ modifies both transport coefficients in a controlled manner:
	\begin{itemize}
		\item The DC conductivity receives a correction proportional to $q_2$ (with $q_1$ set to $1$):
		\begin{equation}
			\sigma = 1 - q_2 \left( \frac{9\kappa Q^2}{L^2 r_h^4} + \frac{7\kappa^2 Q^4}{4 r_h^8} \right).
		\end{equation}
		For $q_2>0$, the conductivity decreases below the unit value, thereby violating the quantum critical conductivity bound, while for $q_2<0$ the bound remains satisfied. The bound is saturated exactly when $q_2=0$, recovering the standard result of the minimal theory.
		
		\item The shear viscosity to entropy density ratio is modified as
		\begin{equation}
			\frac{\eta}{s} = \frac{1}{4\pi}\left(1 + q_2\,\frac{7\kappa^2 Q^4}{2 r_h^8}\right).
		\end{equation}
		This shows that the KSS bound $\eta/s \ge 1/(4\pi)$ is not universally preserved. The sign of $q_2$ determines whether the ratio is enhanced ($q_2>0$) or suppressed ($q_2<0$) relative to the Einstein gravity value. In the limit $q_2\to 0$, the standard result is recovered.
	\end{itemize}
	
	These findings demonstrate that the nonminimal coupling $F^{(a)\alpha\beta}F^{(a)\gamma\lambda}R_{\alpha\gamma}R_{\beta\lambda}$ has a significant impact on the hydrodynamic transport of the dual non-Abelian plasma. The sign of the coupling plays a crucial role, with positive $q_2$ leading to a reduction in conductivity and an increase in $\eta/s$, while negative $q_2$ yields opposite effects. This sensitivity to the sign of the higher-derivative coupling is reminiscent of similar phenomena observed in other holographic models with curvature corrections.
	
	Our perturbative analysis provides a consistent framework for exploring the consequences of such nonminimal interactions. The small parameter $q_2$ is assumed to be sufficiently small so that the first-order correction is a reliable approximation.  A natural extension would be to compute higher-order corrections to assess the convergence of the expansion and to determine whether the observed violations of the transport bounds persist or are compensated by higher-order terms.  Additionally, investigating the full frequency-dependent conductivities and the full structure of the hydrodynamic dispersion relations could reveal further signatures of the nonminimal coupling.  From a broader perspective, the model considered here may serve as a holographic laboratory for studying non-Abelian plasmas with curvature-induced modifications, potentially offering insights into the behavior of strongly coupled gauge theories in extreme environments such as the quark-gluon plasma.
	
	We also note several directions suggested by the referee comments. First, we have restricted ourselves to an electric ansatz for the Yang-Mills potential. A dyonic ansatz (including magnetic charge) would likely introduce additional corrections to the transport coefficients proportional to the magnetic charge, and it would be interesting to study such a generalization in future work. Second, similar results are expected for a nonminimal $U(1)$ electromagnetic theory, since the essential structure of the coupling is analogous. Third, the nonminimal term considered here also appears in the Kaluza-Klein reduction of higher-dimensional Euler forms, as shown in \cite{Dereli1990}. Fourth, an alternative approach using the first-order formalism (vielbein and spin connection) would allow for propagating torsion; examining the effect of torsion on the hydrodynamic coefficients is an intriguing open problem. Finally, extending the analysis to even higher-order couplings such as $R F^4$ would be a natural step towards a more complete effective field theory treatment.
	
	%--------------------------------------------
	\vspace{1cm}
\noindent {\large {\bf Acknowledgment} }\\\\
We also extend our sincere thanks to the referees of Classical and Quantum Gravity for their valuable comments, which have enabled us to enhance the quality of the manuscript.

\vspace{1cm}
\noindent {\large {\bf Data Availability} } No new data were created or analysed in this study.

	%%%%%%%%%%%%%%%%%%%%%%%%%%%%%%%%%%%%%%%%%%%%%%%%%%%
	
	\appendix
	\section{Explicit form of \(T^{(I)}_{\mu\nu}\)}
	\label{app:A}
	The energy-momentum tensor arising from the non-minimal coupling is as,
	\begin{eqnarray}
		T^{(I)}_{\mu \nu }&=&\tfrac{1}{2} F^{(a) \alpha \beta } F^{(a)\gamma \lambda } g_{\mu \nu } R_{\alpha \gamma } R_{\beta \lambda } - 2 F^{(a)\beta \gamma } F_{\nu }{}^{(a)\alpha } R_{\alpha \gamma } R_{\mu \beta } \nonumber \\ 
		&& - 2 F^{(a)\beta \gamma } F_{\mu }{}^{(a)\alpha } R_{\alpha \gamma } R_{\nu \beta } -  F^{(a)\alpha \beta } g_{\mu \nu } R_{\alpha \gamma } \nabla_{\beta }\nabla_{\lambda }F^{(a)\gamma \lambda } \nonumber \\ 
		&& -  F^{(a)\alpha \beta } R_{\alpha \gamma } \nabla_{\beta }\nabla_{\mu }F_{\nu }{}^{(a)\gamma } -  F^{(a)\alpha \beta } R_{\alpha \gamma } \nabla_{\beta }\nabla_{\nu }F_{\mu }{}^{(a)\gamma } \nonumber \\ 
		&& + 2 F^{(a)\alpha \beta } g_{\mu \nu } \nabla_{\beta }R_{\alpha }{}^{\lambda } \nabla_{\gamma }F^{(a)\gamma }{}_{\lambda } -  F_{\nu }{}^{(a)\alpha } R_{\alpha \beta } \nabla_{\gamma }\nabla^{\gamma }F_{\mu }{}^{(a)\beta } \nonumber \\ 
		&& -  F_{\mu }{}^{(a)\alpha } R_{\alpha \beta } \nabla_{\gamma }\nabla^{\gamma }F_{\nu }{}^{(a)\beta } -  F_{\mu }{}^{(a)\alpha } F_{\nu }{}^{(a)\beta } \nabla_{\gamma }\nabla^{\gamma }R_{\alpha \beta } \nonumber \\ 
		&& -  F_{\nu }{}^{(a)\alpha } R_{\alpha \beta } \nabla_{\gamma }\nabla_{\mu }F^{(a)\beta \gamma } -  F^{(a)\beta \gamma } F_{\nu }{}^{(a)\alpha } \nabla_{\gamma }\nabla_{\mu }R_{\alpha \beta } \nonumber \\ 
		&& -  F_{\mu }{}^{(a)\alpha } R_{\alpha \beta } \nabla_{\gamma }\nabla_{\nu }F^{(a)\beta \gamma } -  F^{(a)\beta \gamma } F_{\mu }{}^{(a)\alpha } \nabla_{\gamma }\nabla_{\nu }R_{\alpha \beta } \nonumber \\ 
		&& - 2 R_{\alpha \beta } \nabla_{\gamma }F_{\mu }{}^{(a)\alpha } \nabla^{\gamma }F_{\nu }{}^{(a)\beta } - 2 F_{\nu }{}^{(a)\alpha } \nabla_{\gamma }F_{\mu }{}^{(a)\beta } \nabla^{\gamma }R_{\alpha \beta } \nonumber \\ 
		&& - 2 F_{\mu }{}^{(a)\alpha } \nabla_{\gamma }F_{\nu }{}^{(a)\beta } \nabla^{\gamma }R_{\alpha \beta } -  g_{\mu \nu } R_{\alpha \beta } \nabla_{\gamma }F^{(a)\beta \lambda } \nabla_{\lambda }F^{(a)\alpha \gamma } \nonumber \\ 
		&& -  g_{\mu \nu } R_{\alpha \beta } \nabla_{\gamma }F^{(a)\alpha \gamma } \nabla_{\lambda }F^{(a)\beta \lambda } + 2 F^{(a)\alpha \beta } g_{\mu \nu } \nabla_{\alpha }F^{(a)\gamma \lambda } \nabla_{\lambda }R_{\beta \gamma } \nonumber \\ 
		&& -  F^{(a)\alpha \beta } g_{\mu \nu } R_{\alpha \gamma } \nabla_{\lambda }\nabla_{\beta }F^{\gamma \lambda } -  F^{(a)\alpha \beta } F^{(a)\gamma \lambda } g_{\mu \nu } \nabla_{\lambda }\nabla_{\beta }R_{\alpha \gamma } \nonumber \\ 
		&& -  F_{\nu }{}^{(a)\alpha } \nabla_{\gamma }R_{\alpha }{}^{\beta } \nabla_{\mu }F^{(a)}_{\beta }{}^{\gamma } -  R_{\alpha \beta } \nabla_{\gamma }F_{\nu }{}^{(a)\alpha } \nabla_{\mu }F^{(a)\beta \gamma } \nonumber \\ 
		&& -  R_{\alpha \beta } \nabla_{\gamma }F^{(a)\beta \gamma } \nabla_{\mu }F_{\nu }{}^{(a)\alpha } -  F^{(a)\alpha \beta } \nabla_{\beta }R_{\alpha }{}^{\gamma } \nabla_{\mu }F^{(a)}_{\nu \gamma } \nonumber \\ 
		&& + F_{\nu }{}^{(a)\alpha } \nabla_{\beta }F^{(a)\beta \gamma } \nabla_{\mu }R_{\alpha \gamma } + F^{(a)\alpha \beta } \nabla_{\alpha }F_{\nu }{}^{(a)\gamma } \nabla_{\mu }R_{\beta \gamma } \nonumber \\ 
		&& -  F_{\mu }{}^{(a)\alpha } \nabla_{\gamma }R_{\alpha }{}^{\beta } \nabla_{\nu }F_{\beta }{}^{(a)\gamma } -  R_{\alpha \beta } \nabla_{\gamma }F_{\mu }{}^{(a)\alpha } \nabla_{\nu }F^{(a)\beta \gamma } \nonumber \\ 
		&& -  R_{\alpha \beta } \nabla_{\gamma }F^{(a)\beta \gamma } \nabla_{\nu }F_{\mu }{}^{(a)\alpha } -  F^{(a)\alpha \beta } \nabla_{\beta }R_{\alpha }{}^{\gamma } \nabla_{\nu }F^{(a)}_{\mu \gamma } \nonumber \\ 
		&& + F_{\mu }{}^{(a)\alpha } \nabla_{\beta }F^{(a)\beta \gamma } \nabla_{\nu }R_{\alpha \gamma } + F^{(a)\alpha \beta } \nabla_{\alpha }F^{(a)}_{\mu }{}^{\gamma } \nabla_{\nu }R_{\beta \gamma }.
	\end{eqnarray}
	This expression is used in the derivation of the gravitational field equations.
	
	\section{Explicit expressions for \(B_1(r), B_2(r), B_3(r)\)}
	\label{app:B}
	The functions \(B_1(r)\), \(B_2(r)\), and \(B_3(r)\) that appear in the \(tt\), \(rr\), and \(xx\) components of the gravitational equations are,
	\begin{align}
		B_1(r) &=  8 f'(r)^2 h'(r)^2 - 28 r f'(r)^2 h'(r)^2 H'(r) + 24 f(r)^2 h'(r)^2 H'(r)^2 \nonumber\\
		&\quad + 108 f(r) r f'(r) h'(r)^2 H'(r)^2 + 35 r^2 f'(r)^2 h'(r)^2 H'(r)^2 - 40 f(r)^2 r h'(r)^2 H'(r)^3 \nonumber\\
		&\quad - 28 f(r) r^2 f'(r) h'(r)^2 H'(r)^3 - 20 f(r)^2 r^2 h'(r)^2 H'(r)^4 - 12 f(r) h'(r)^2 f''(r) \nonumber\\
		&\quad + 4 r f'(r) h'(r)^2 f''(r) - 24 f(r) r h'(r)^2 H'(r) f''(r) - 16 r^2 f'(r) h'(r)^2 H'(r) f''(r) \nonumber\\
		&\quad + 36 f(r) r^2 h'(r)^2 H'(r)^2 f''(r) + 3 r^2 h'(r)^2 f''(r)^2 - 16 f(r) f'(r) h'(r) h''(r) \nonumber\\
		&\quad - 8 r f'(r)^2 h'(r) h''(r) + 32 f(r)^2 h'(r) H'(r) h''(r) + 48 f(r) r f'(r) h'(r) H'(r) h''(r) \nonumber\\
		&\quad + 12 r^2 f'(r)^2 h'(r) H'(r) h''(r) + 16 f(r)^2 r h'(r) H'(r)^2 h''(r) + 56 f(r) r^2 f'(r) h'(r) H'(r)^2 h''(r) \nonumber\\
		&\quad - 64 f(r)^2 r^2 h'(r) H'(r)^3 h''(r) - 56 f(r) r h'(r) f''(r) h''(r) - 4 r^2 f'(r) h'(r) f''(r) h''(r) \nonumber\\
		&\quad + 16 f(r) r^2 h'(r) H'(r) f''(r) h''(r) - 16 f(r) r f'(r) h''(r)^2 + 16 f(r)^2 r H'(r) h''(r)^2 \nonumber\\
		&\quad + 24 f(r) r^2 f'(r) H'(r) h''(r)^2 - 16 f(r)^2 r^2 H'(r)^2 h''(r)^2 - 8 f(r) r^2 f''(r) h''(r)^2 \nonumber\\
		&\quad + 24 f(r)^2 h'(r)^2 H''(r) + 48 f(r) r f'(r) h'(r)^2 H''(r) + 10 r^2 f'(r)^2 h'(r)^2 H''(r) \nonumber\\
		&\quad + 72 f(r)^2 r h'(r)^2 H'(r) H''(r) + 108 f(r) r^2 f'(r) h'(r)^2 H'(r) H''(r) \nonumber\\
		&\quad - 72 f(r)^2 r^2 h'(r)^2 H'(r)^2 H''(r) + 12 f(r) r^2 h'(r)^2 f''(r) H''(r) \nonumber\\
		&\quad + 80 f(r)^2 r h'(r) h''(r) H''(r) + 88 f(r) r^2 f'(r) h'(r) h''(r) H''(r) \nonumber\\
		&\quad + 16 f(r)^2 r^2 h''(r)^2 H''(r) + 12 f(r)^2 r^2 h'(r)^2 H''(r)^2 - 20 f(r) r h'(r)^2 f^{(3)}(r) \nonumber\\
		&\quad - 2 r^2 f'(r) h'(r)^2 f^{(3)}(r) - 4 f(r) r^2 h'(r)^2 H'(r) f^{(3)}(r) - 16 f(r) r^2 h'(r) h''(r) f^{(3)}(r) \nonumber\\
		&\quad - 16 f(r) r f'(r) h'(r) h^{(3)}(r) + 16 f(r)^2 r h'(r) H'(r) h^{(3)}(r) \nonumber\\
		&\quad + 24 f(r) r^2 f'(r) h'(r) H'(r) h^{(3)}(r) - 16 f(r)^2 r^2 h'(r) H'(r)^2 h^{(3)}(r) \nonumber\\
		&\quad - 8 f(r) r^2 h'(r) f''(r) h^{(3)}(r) + 16 f(r)^2 r^2 h'(r) H''(r) h^{(3)}(r) \nonumber\\
		&\quad + 32 f(r)^2 r h'(r)^2 H^{(3)}(r) + 32 f(r) r^2 f'(r) h'(r)^2 H^{(3)}(r) \nonumber\\
		&\quad + 16 f(r)^2 r^2 h'(r)^2 H'(r) H^{(3)}(r) + 32 f(r)^2 r^2 h'(r) h''(r) H^{(3)}(r) \nonumber\\
		&\quad - 4 f(r) r^2 h'(r)^2 f^{(4)}(r) + 8 f(r)^2 r^2 h'(r)^2 H^{(4)}(r),
	\end{align}
	
	\begin{align}
		B_2(r) &= -8 f'(r)^2 h'(r)^2 - 8 f(r) f'(r) h'(r)^2 H'(r) + 28 r f'(r)^2 h'(r)^2 H'(r) \nonumber\\
		&\quad -60 f(r) r f'(r) h'(r)^2 H'(r)^2 -35 r^2 f'(r)^2 h'(r)^2 H'(r)^2 +40 f(r)^2 r h'(r)^2 H'(r)^3 \nonumber\\
		&\quad +60 f(r) r^2 f'(r) h'(r)^2 H'(r)^3 -28 f(r)^2 r^2 h'(r)^2 H'(r)^4 +12 f(r) h'(r)^2 f''(r) \nonumber\\
		&\quad -4 r f'(r) h'(r)^2 f''(r) -8 f(r) r h'(r)^2 H'(r) f''(r) +16 r^2 f'(r) h'(r)^2 H'(r) f''(r) \nonumber\\
		&\quad -8 f(r) r^2 h'(r)^2 H'(r)^2 f''(r) -3 r^2 h'(r)^2 f''(r)^2 +16 f(r) f'(r) h'(r) h''(r) \nonumber\\
		&\quad +8 r f'(r)^2 h'(r) h''(r) -32 f(r)^2 h'(r) H'(r) h''(r) -48 f(r) r f'(r) h'(r) H'(r) h''(r) \nonumber\\
		&\quad -12 r^2 f'(r)^2 h'(r) H'(r) h''(r) +32 f(r)^2 r h'(r) H'(r)^2 h''(r) +32 f(r) r^2 f'(r) h'(r) H'(r)^2 h''(r) \nonumber\\
		&\quad -16 f(r)^2 r^2 h'(r) H'(r)^3 h''(r) +8 f(r) r h'(r) f''(r) h''(r) +4 r^2 f'(r) h'(r) f''(r) h''(r) \nonumber\\
		&\quad -8 f(r) r^2 h'(r) H'(r) f''(r) h''(r) -24 f(r)^2 h'(r)^2 H''(r) -8 f(r) r f'(r) h'(r)^2 H''(r) \nonumber\\
		&\quad -10 r^2 f'(r)^2 h'(r)^2 H''(r) -16 f(r) r^2 f'(r) h'(r)^2 H'(r) H''(r) +24 f(r)^2 r^2 h'(r)^2 H'(r)^2 H''(r) \nonumber\\
		&\quad +12 f(r) r^2 h'(r)^2 f''(r) H''(r) -16 f(r)^2 r h'(r) h''(r) H''(r) -8 f(r) r^2 f'(r) h'(r) h''(r) H''(r) \nonumber\\
		&\quad +16 f(r)^2 r^2 h'(r) H'(r) h''(r) H''(r) -12 f(r)^2 r^2 h'(r)^2 H''(r)^2 +4 f(r) r h'(r)^2 f^{(3)}(r) \nonumber\\
		&\quad +2 r^2 f'(r) h'(r)^2 f^{(3)}(r) -4 f(r) r^2 h'(r)^2 H'(r) f^{(3)}(r) -8 f(r)^2 r h'(r)^2 H^{(3)}(r) \nonumber\\
		&\quad -4 f(r) r^2 f'(r) h'(r)^2 H^{(3)}(r) +8 f(r)^2 r^2 h'(r)^2 H'(r) H^{(3)}(r),
	\end{align}
	
	\begin{align}
		B_3(r) &= -16 f'(r)^2 h'(r)^2 H'(r) - 16 f(r) f'(r) h'(r)^2 H'(r)^2 \nonumber\\
		&\quad + r f'(r)^2 h'(r)^2 H'(r)^2 - 4 f(r) r f'(r) h'(r)^2 H'(r)^3 + 12 f(r)^2 r h'(r)^2 H'(r)^4 \nonumber\\
		&\quad + 8 f'(r) h'(r)^2 f''(r) - 8 f(r) h'(r)^2 H'(r) f''(r) - 8 r f'(r) h'(r)^2 H'(r) f''(r) \nonumber\\
		&\quad - 6 f(r) r h'(r)^2 H'(r)^2 f''(r) + r h'(r)^2 f''(r)^2 + 8 f'(r)^2 h'(r) h''(r) \nonumber\\
		&\quad - 56 f(r) f'(r) h'(r) H'(r) h''(r) - 12 r f'(r)^2 h'(r) H'(r) h''(r) \nonumber\\
		&\quad - 16 f(r)^2 h'(r) H'(r)^2 h''(r) - 12 f(r) r f'(r) h'(r) H'(r)^2 h''(r) \nonumber\\
		&\quad + 24 f(r)^2 r h'(r) H'(r)^3 h''(r) + 24 f(r) h'(r) f''(r) h''(r) + 4 r f'(r) h'(r) f''(r) h''(r) \nonumber\\
		&\quad - 12 f(r) r h'(r) H'(r) f''(r) h''(r) + 8 f(r) f'(r) h''(r)^2 - 16 f(r)^2 H'(r) h''(r)^2 \nonumber\\
		&\quad - 12 f(r) r f'(r) H'(r) h''(r)^2 + 8 f(r)^2 r H'(r)^2 h''(r)^2 + 4 f(r) r f''(r) h''(r)^2 \nonumber\\
		&\quad - 48 f(r) f'(r) h'(r)^2 H''(r) - 10 r f'(r)^2 h'(r)^2 H''(r) - 16 f(r)^2 h'(r)^2 H'(r) H''(r) \nonumber\\
		&\quad - 14 f(r) r f'(r) h'(r)^2 H'(r) H''(r) + 16 f(r)^2 r h'(r)^2 H'(r)^2 H''(r) \nonumber\\
		&\quad - 16 f(r) r h'(r)^2 f''(r) H''(r) - 48 f(r)^2 h'(r) h''(r) H''(r) \nonumber\\
		&\quad - 48 f(r) r f'(r) h'(r) h''(r) H''(r) + 8 f(r)^2 r h'(r) H'(r) h''(r) H''(r) \nonumber\\
		&\quad - 8 f(r)^2 r h''(r)^2 H''(r) + 4 f(r)^2 r h'(r)^2 H''(r)^2 \nonumber\\
		&\quad + 8 f(r) h'(r)^2 f^{(3)}(r) + 2 r f'(r) h'(r)^2 f^{(3)}(r) + 8 f(r) r h'(r) h''(r) f^{(3)}(r) \nonumber\\
		&\quad + 8 f(r) f'(r) h'(r) h^{(3)}(r) - 16 f(r)^2 h'(r) H'(r) h^{(3)}(r) \nonumber\\
		&\quad - 12 f(r) r f'(r) h'(r) H'(r) h^{(3)}(r) + 8 f(r)^2 r h'(r) H'(r)^2 h^{(3)}(r) \nonumber\\
		&\quad + 4 f(r) r h'(r) f''(r) h^{(3)}(r) - 8 f(r)^2 r h'(r) H''(r) h^{(3)}(r) \nonumber\\
		&\quad - 16 f(r)^2 h'(r)^2 H^{(3)}(r) - 18 f(r) r f'(r) h'(r)^2 H^{(3)}(r) \nonumber\\
		&\quad - 4 f(r)^2 r h'(r)^2 H'(r) H^{(3)}(r) - 16 f(r)^2 r h'(r) h''(r) H^{(3)}(r) \nonumber\\
		&\quad + 2 f(r) r h'(r)^2 f^{(4)}(r) - 4 f(r)^2 r h'(r)^2 H^{(4)}(r),
	\end{align}
	These expressions are used in the derivation of the perturbative solution.

	%--------------------------------------------------------------------------

\end{document}